\documentclass[10pt,letterpaper,twocolumn]{article} 

\usepackage{ol2}
\usepackage{mathptmx,amsmath,amssymb}
\usepackage{graphicx}

\graphicspath{{pictOL/}{pict/}{}}

\newcommand\pictc[5]{\begin{figure}
                   \centerline{
                   \includegraphics[width=#1\columnwidth,height=0.8\textheight,keepaspectratio]{#3}}
               \protect\caption{\protect\label{fig:#4} #5}
                \end{figure}            }

\newcommand\pict[4][1]{\pictc{#1}{!tb}{#2}{#3}{#4}}

\newcommand\rpict[1]{\ref{fig:#1}}

\newcommand\leqt[1]{\protect\label{eq:#1}}
\newcommand\reqtn[1]{\ref{eq:#1}}
\newcommand\reqt[1]{(\reqtn{#1})}

\newcounter{Fig}

\begin{document}
\twocolumn[ 

\title{Slow-light enhanced optical forces between \\ longitudinally shifted photonic-crystal nanowire waveguides}

\author{Yue Sun$^{1,2,\ast}$, Thomas P. White$^{1,2}$, and Andrey A. Sukhorukov$^{1}$}
\address{$^1$ Nonlinear Physics Centre and $^2$ Laser Physics Centre, Centre for Ultrahigh-bandwidth Devices for Optical Systems, \\
Research School of Physics and Engineering, Australian National University, Canberra
ACT 0200, Australia\\
$^\ast$Corresponding author: sue124@physics.anu.edu.au}

\begin{abstract}
We reveal that slow-light enhanced optical forces between side-coupled photonic-crystal nanowire waveguides can be flexibly controlled by introducing a relative longitudinal shift. We predict that close to the photonic band-edge, where the group velocity is reduced, the transverse force can be tuned from repulse to attractive, and the force is suppressed for a particular shift value. Additionally the shift leads to symmetry breaking that can facilitate longitudinal forces acting on the waveguides, in contrast to unshifted structures where such forces vanish.
\end{abstract}


] 


Development of reconfigurable photonic circuits utilizing gradient optical forces between micro- and nano-scale optical waveguides opens new possibilities for optical signal shaping and routing based on all-optical tuning of the structure geometry~\cite{Li:2009-464:NPHOT, Roels:2009-510:NNANO, Eichenfield:2009-550:NAT}. Optomechanical interactions can be enhanced in photonic-crystal waveguides due to increased optical energy concentration in the slow-light regime~\cite{Ma:2010-151102:APL} or through light trapping in cavities~\cite{Chan:2009-3802:OE}. Whereas the pump intensity determines the strength of gradient force, the sign of the gradient force between two identical coupled waveguides or cavities is dependant on the modal symmetry, with even modes typically associated with attractive and odd modes with repulsive forces~\cite{Povinelli:2005-3042:OL}. This has been demonstrated in side-coupled waveguides, where the force can be tuned from attractive to repulsive by varying the phase of the incident light to excite modes of different symmetries~\cite{Li:2009-464:NPHOT, Roels:2009-510:NNANO}.

In this work, we suggest and demonstrate through 3D numerical simulations that slow-light enhanced optically induced forces between side-coupled photonic-crystal wire waveguides can be tuned by introducing a relative longitudinal shift. This happens because the shift breaks the structure symmetry, which leads to pronounced hybridization of odd and even modes at the resonant frequencies close to the photonic band-edge where the group velocity is reduced. This enables the tuning of the transverse force from repulse to attractive, and force suppression for a particular shift value.
We also reveal
that the symmetry breaking can facilitate longitudinal forces acting on the waveguides, in contrast to unshifted structures where such forces vanish.


\pict{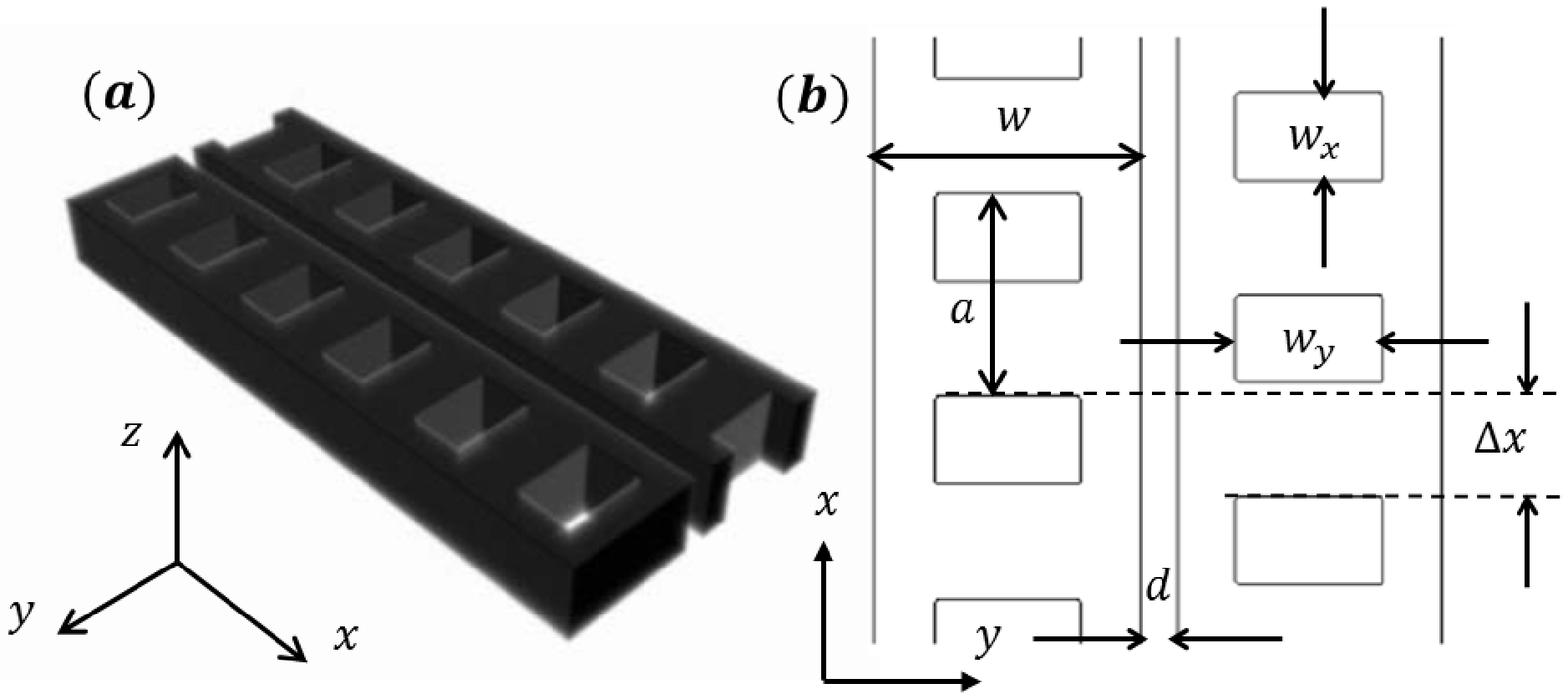}{structure}{
(a)~3D sketch of the longitudinally shifted side-coupled photonic-crystal waveguides.
(b)~top view of the structure.
}

We analyze suspended wire waveguides containing a periodic array of rectangular holes as illustrated in Fig.~\rpict{structure}(a), and investigate the effect of the longitudinal shift $\Delta x$.
To provide a link with experimental platform, we consider silicon nitride material (refractive index $n \simeq 2$) and choose the dimensions according to Ref.~\cite{Chan:2009-3802:OE}, but without a cavity since we aim to utilize slow-light enhancement of interactions which can be realized in defect-free photonic-crystal waveguides~\cite{Ma:2010-151102:APL}.
The structure sketch is presented in Fig.~\rpict{structure}. The period of the rectangular air hole array is $a$.
The width and height of both waveguides are $w={7}a/{6}$ and $h={2}a/{3}$, respectively, and the space between them is $d=a/{6}$. The width and length of the rectangular air holes are $w_x=0.445 a$ and $w_y={2}a/{3}$ respectively.

\pict{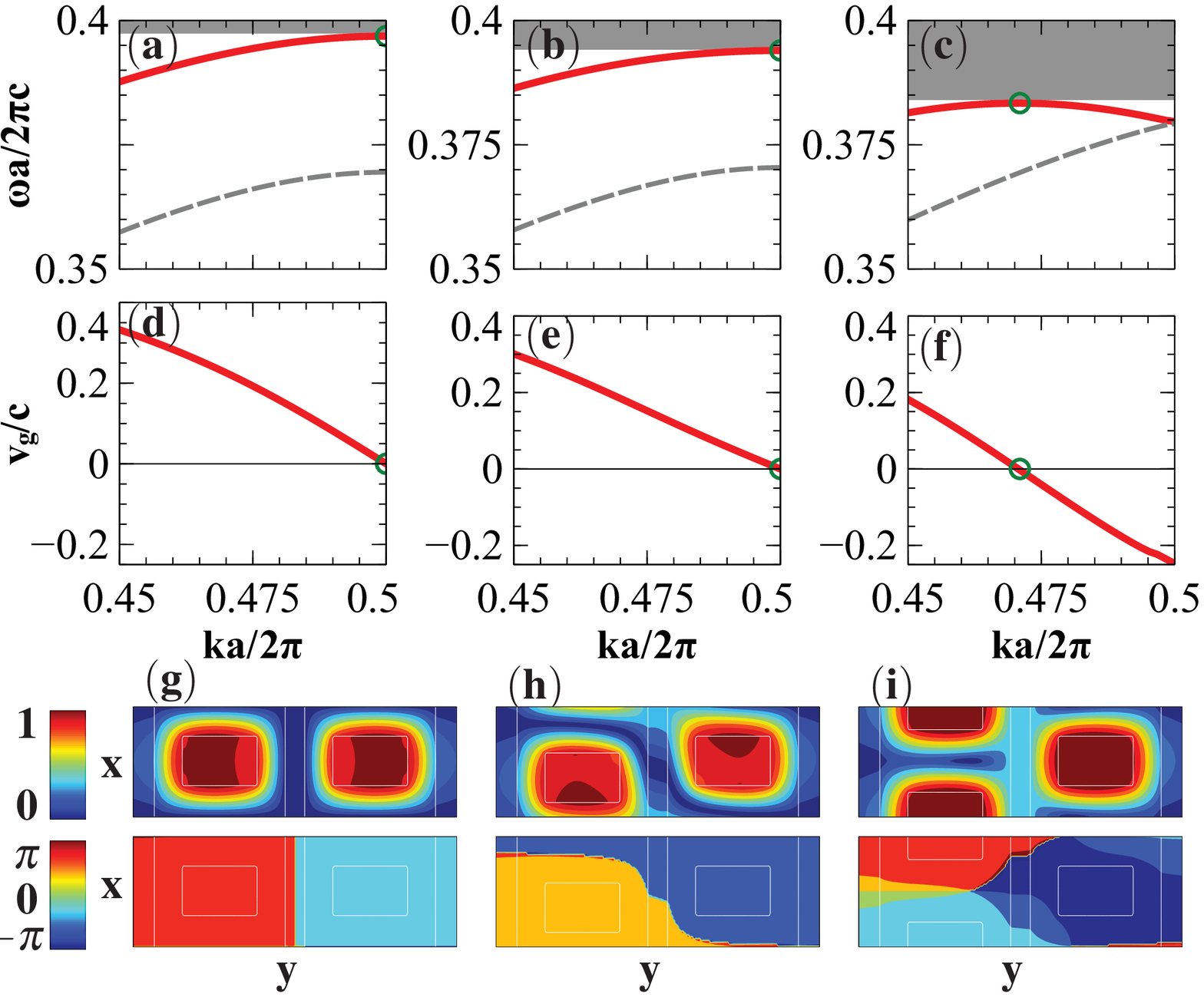}{dispersion}{
(Color online) TM mode properties in coupled waveguides with different longitudinal shifts:
(a,d,g)~$\Delta x=0$, (b,e,h)~$\Delta x = 0.15 a$, and (c,f,i)~$\Delta x = 0.5 a$.
(a,b,c)~Dispersion relations shown as normalized frequency vs. wavenumber for the first- (dashed) and second-lowest (solid line) TM bands near the band-edge; grey shading marks the band-gap. (d,e,f)~Group velocities for the second-lowest band.
(g,h,i)~Mode profiles at the band-edge: intensity $|E|^2$ (top row) and phase (bottom row), white contours show an outline of the waveguides.
}

We use a plane-wave expansion method~\cite{Johnson:2001-173:OE} to calculate the supermode dispersion and the full 3D vectorial electromagnetic fields.
In Figs.~\rpict{dispersion}(a-c), we plot the dispersion curves for the first two transverse-magnetic (TM) bands close to the photonic band-edge, where the second band is the one we consider in the remainder of this paper. The corresponding group velocities for the second band are shown in Figs.~\rpict{dispersion}(d-f). The plots are presented in dimensionless units: normalized frequency $\omega a / 2 \pi c$, normalized group velocity $v_g / c$, and normalized wavenumber $k a / 2 \pi$, where $\omega$ is optical angular frequency, $v_g$ is the group velocity, $k$ is the wavevector component along the waveguides, and $c$ is the speed of light in vacuum.
The results are shown for three values of the longitudinal shift, as indicated in the figure caption.
The group velocity vanishes at the photonic band-edge for all shift values, and this is a generally known feature.
Additionally, it is also widely accepted that light-matter interactions can be increased in the slow-light regime, including an enhancement of the optical gradient forces~\cite{Ma:2010-151102:APL}. However, as we demonstrate below, the introduction of a longitudinal shift has a nontrivial effect on slow-light enhanced optical forces, which can be attributed to breaking of symmetry of band-edge modes from odd (anti-symmetric) profiles in the absence of a shift [Figs.~\rpict{dispersion}(g)], to complex phase profiles for shifted waveguides [Figs.~\rpict{dispersion}(h,i)].

We use the profiles of the optical modes to calculate the time-averaged gradient forces acting on the waveguides. We take into account the structure periodicity and associated Bloch symmetry properties of electromagnetic modes to simplify the general expressions based on the Maxwell stress tensor formulation $T_{\alpha \beta}$~\cite{Jackson:1998:ClassicalElectrodynamics}. After excluding terms with vanishing contribution, we obtain the following expressions for the distributed forces acting on
the right waveguide in the transverse ($y$-axis) and longitudinal ($x$-axis) directions,
\begin{equation} \leqt{Fdistrib}
  f_y = -  \int_{-\infty}^{+\infty} dz T_{yy}|_{y=0}, \quad
  f_x = -  \int_{-\infty}^{+\infty} dz T_{xy}|_{y=0} .
\end{equation}
In these expressions, the coordinate axis origin $y=0$ lies in the middle between the nanowire waveguides, and at this cross-section the tensor components are defined as
\begin{align}
    \leqt{Tyy}
    T_{yy}|_{y=0} &= |E_y|^2-|E_x|^2-|E_z|^2 + |H_y|^2-|H_x|^2-|H_z|^2 , \\
    \leqt{Txy}
    T_{xy}|_{y=0} &= E_x^\ast E_y + E_x E_y^\ast + H_x^\ast H_y + H_x H_y^\ast.
\end{align}
The force in the vertical direction vanishes, $f_z=0$, if the nanowires are free-standing or suspended high above a substrate.
The gradient forces for the left waveguide have opposite values to those for the right waveguide given in Eq.~\reqt{Fdistrib}.
We also analyze the net forces over one period, calculated as
\begin{equation}\label{Fperiod}
  F_{x,y} = \int_{-{a}/{2}}^{{a}/{2}} f_{x,y} dx .
\end{equation}

For comparison between different structures, it is convenient to consider a normalized dimensionless force~\cite{Povinelli:2005-3042:OL}, $\widetilde{F} = F c / P$, where $P$ is the power flow through the waveguides.
In the absence of losses the optical energy stored within one period of the structure can be expressed as $U = P a / |v_g|$. As the group velocity is reduced close to the photonic band-edge, the value of $U$ increases, and the force value also grows as $\widetilde{F} \sim |v_g|^{-1}$. In order to compare forces in the slow-light regime for different structure parameters, we also consider a scaled force value $\widetilde{F} |v_g|/ c \equiv F |v_g| / P$.

\pict{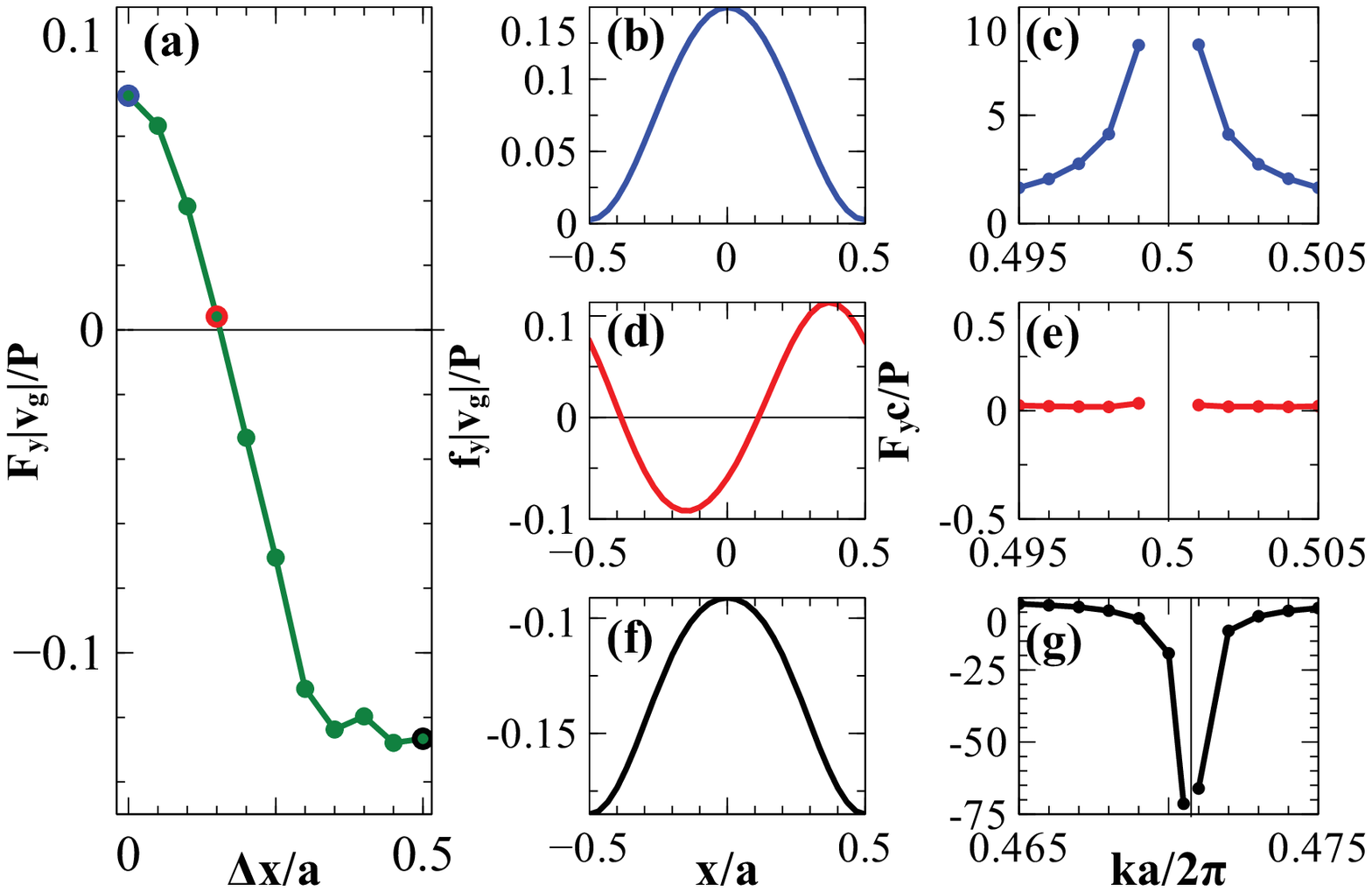}{Forcey}{
The force in the transverse ($+y$) direction on the right waveguide.
(a)~The force per unit length per unit energy density vs. the longitudinal shift at the band-edge.
\mbox{(b-g)}~Forces for different longitudinal shifts: (b,c)~$\Delta x=0$, (d,e)~$\Delta x=0.15 a$ and (f,g)~$\Delta x=0.5 a$.
(b,d,f)~The distributed force along the waveguide per unit energy density at the band-edge. (c,e,g)~Total force per unit length and unit power vs. the wavenumber near the photonic band-edge.
}

In the transverse direction, the optically induced force originates from the strong variation of the electromagnetic field due to the evanescent coupling between two waveguides. Most importantly, we show in Fig.~\rpict{Forcey}(a) that the slow-light enhanced force at the photonic band edge changes from repulsive (positive value) to attractive (negative value) as the longitudinal shift is tuned from zero ($\Delta x=0$) to half a period ($\Delta x = 0.5 a$).
In the absence of a longitudinal shift ($\Delta x = 0$), the band-edge mode has odd symmetry about $y$-axis [Fig.~\rpict{dispersion}(g)] which provides a repulsive force at all waveguide locations [Fig.~\rpict{Forcey}(b)] in agreement with earlier studies for non-periodic waveguides~\cite{Povinelli:2005-3042:OL}, and in periodic waveguides the force is additionally enhanced close to photonic band-edge, see Fig.~\rpict{Forcey}(c).
When a longitudinal shift is introduced, the left-right symmetry of the coupled waveguide structure is broken, and this leads to cross-coupling between odd and even supermodes.
Consequently, the electromagnetic fields become a mixture of even and odd modes [Fig.~\rpict{dispersion}(h,i)] and the distributed force along the $x$ direction changes its offset and peak-to-peak value, c.f. Figs.~\rpict{Forcey}(b,d,f).
As the longitudinal shift increases, the relative fraction of even supermodes gets larger, which
explains the transition to transversal attraction between waveguides at larger shifts. In particular, the strongest slow-light enhanced attraction occurs for half a period shift ($\Delta x = 0.5 a$), when the distributed force is attractive at all waveguide locations, see Figs.~\rpict{Forcey}(f,g).
Quite interestingly, for an intermediate shift value of $\Delta x \simeq 0.15 a$, the transverse force is suppressed even in the slow-light regime, see Fig.~\rpict{Forcey}(e), which happens because the distributed force oscillates between repulsive and attractive along the waveguide [Fig.~\rpict{Forcey}(d)] providing zero average force along a structure period.

We estimate the characteristic values of the transverse optical forces, considering $100$~mW optical power coupled into the waveguides and $1550$~nm band-edge wavelength. Then, the transverse force per unit length is 4.45~nN/$\mu$m for $\Delta x = 0$, $a \simeq 617$nm, $ka/2\pi=0.499$, and $v_g/c=0.0081$ [Fig.~\rpict{Forcey}(c)]; 0.02~nN/$\mu$m for $\Delta x = 0.15 a$, $a \simeq 612$nm, $ka/2\pi=0.499$, and $v_g/c=0.0057$ [Fig.~\rpict{Forcey}(e)]; and $-39.91$~nN/${\mu}$m for $\Delta x = 0.5 a$, $a \simeq 596$nm, $ka/2{\pi}=0.4705$, and $v_g/c=0.0018$ [Fig.~\rpict{Forcey}(g)].

\pict{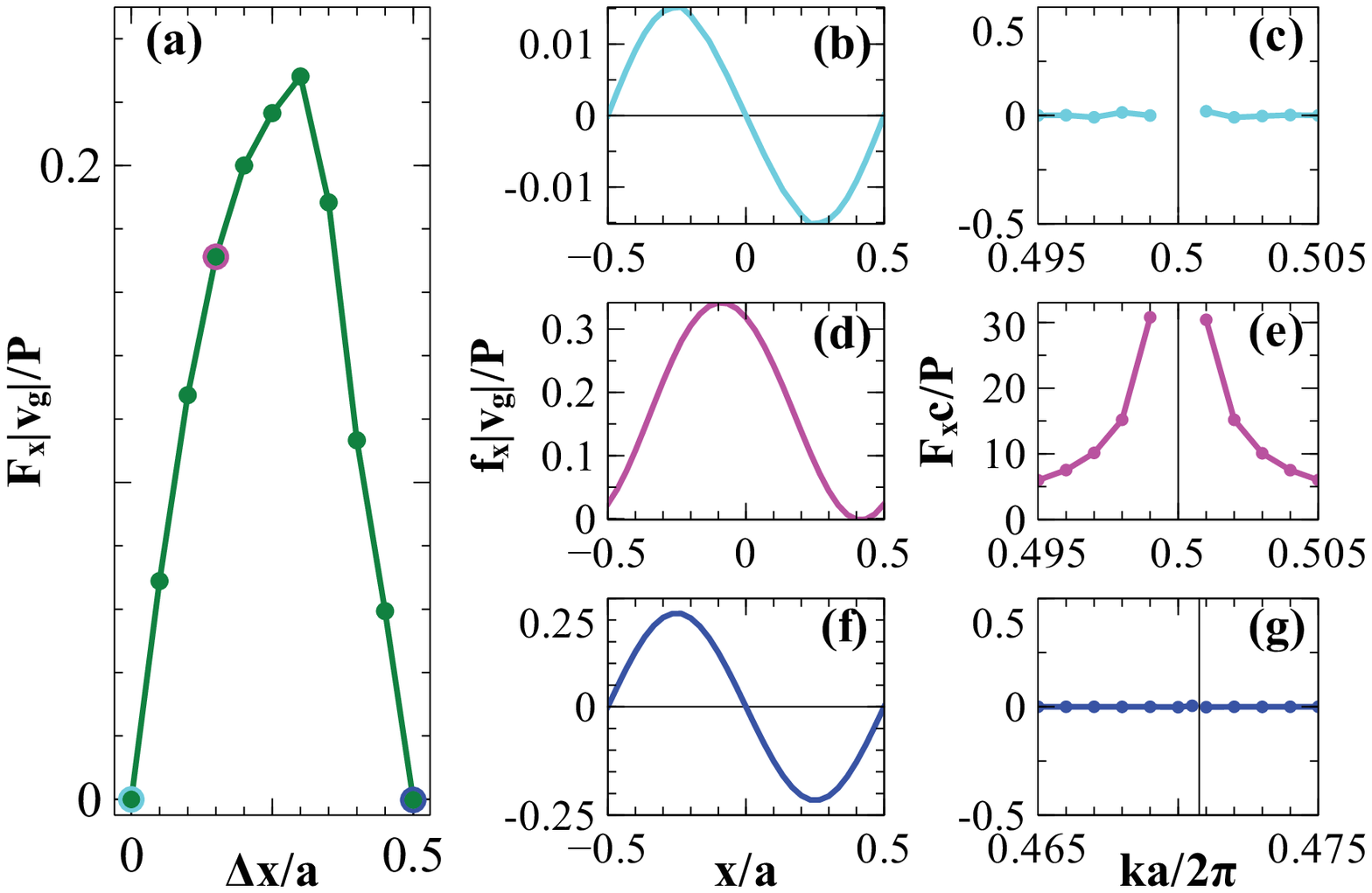}{Forcex}{
The force in the longitudinal ($+x$) direction on the right waveguide.
(a)~The force per unit length per unit energy density vs. the longitudinal shift at the band-edge.
\mbox{(b-g)}~Forces for different longitudinal shifts: (b,c)~$\Delta x=0$, (d,e)~$\Delta x=0.15 a$ and (f,g)~$\Delta x=0.5 a$.
(b,d,f)~The distributed force along the waveguide per unit energy density at the band-edge. (c,e,g)~Total force per unit length and unit power vs. the wavenumber near the photonic band-edge.
}

In the longitudinal direction, the optical force also results from the strong electromagnetic field variation along the $y$ direction according to Eq.~\reqt{Fdistrib},\reqt{Txy}. The scaled force per unit length per unit energy density is shown in Fig.~\rpict{Forcex}(a). We note that it has a maximum value at $\Delta x \simeq 0.3 a$ shift, while zero and half-period shifts correspond to zero force.
In order to explain this force dependence, we use the general symmetry properties of electromagnetic fields in periodic photonic structures.

For zero or half-period shifts, the refractive index profile has a mirror symmetry with respect to a reflection transformation $x \rightarrow -x$, and then it follows~\cite{Joannopoulos:2008:PhotonicCrystals} that Bloch wave profiles should satisfy relations $E_{y,z}(x,y,z; k) = E_{y,z}(-x,y,z; -k) \exp(i \varphi_1)$, $H_{y,z}(x,y,z; k) = - H_{y,z}(-x,y,z; -k) \exp(i \varphi_1)$, $E_{x}(x,y,z; k) = - E_{x}(-x,y,z; -k) \exp(i \varphi_1)$, and $H_{x}(x,y,z; k) = H_{x}(-x,y,z; -k) \exp(i \varphi_1)$. Additionally, for lossless structures with real refractive index profile, there is a symmetry related to time-reversal invariance~\cite{Joannopoulos:2008:PhotonicCrystals}: $\{E,H\}_{x,y,z}^\ast(x,y,z; k) = \{E,H\}_{x,y,z}(x,y,z; -k) \exp(i \varphi_2)$. In these expressions the phase parameters $\varphi_{1,2}$ can take $0$ or $\pi$ values depending on the mode symmetry. Using these symmetry relations, the tensor component responsible for the longitudinal force according to Eqs.~\reqt{Fdistrib},\reqt{Txy} can be transformed as: $T_{xy} =  [- E_x(-x,y,z; k) E_y(x,y,z; k) + E_x(x,y,z; k) E_y(-x,y,z; k) + H_x(-x,y,z; k) H_y(x,y,z; k) - H_x(x,y,z; k) H_y(-x,y,z; k)] \exp(i \varphi)$ where $\varphi = \varphi_1 + \varphi_2$. It is then clear that $T_{xy}(x,y,z;k) =  - T_{xy}(-x,y,z;k)$, and accordingly the distributed longitudinal force is also an antisymmetric function of $x$, $f_{x}(x;k) =  - f_{x}(-x;k)$. Indeed, the antisymmetric property of the distributed force is clearly visible in Figs.~\rpict{Forcex}(b,f). Therefore, for structures with zero or half-period shifts the integral force over one period will always vanish, $F_x \equiv 0$, see the force dependence near the band-edge in Figs.~\rpict{Forcex}(c,g).

For intermediate shift values, $0 < \Delta x < 0.5 a$, the structure's mirror symmetry is broken, resulting in a non-vanishing longitudinal force, see Figs.~\rpict{Forcex}(d,e).
For $100$~mW optical power and $1550$~nm band-edge wavelength, we estimate the force per unit length is
$16.76 nN/{\mu}m$ for $\Delta x = 0.15 a$, $a \simeq 612$nm, $ka/2{\pi}=0.499$ and $v_g/c=0.0057$.

In conclusion, we reveal that slow-light enhanced optical forces between side-coupled photonic-crystal waveguides can be controlled by a longitudinal shift. The shift modifies the overall structure symmetry, enabling force control beyond the possibilities of symmetric structures with unshifted waveguides. In the transverse direction, the longitudinal shift can be used to tune the optical gradient force from repulsive to attractive, and the force can be also suppressed.
The shift also gives rise to force in the longitudinal direction along the waveguide, whereas this force component is always absent in symmetric structures.
These results demonstrate that longitudinal waveguide shifting presents new opportunities for controlling gradient forces which may contribute to the development of optomechanical circuits.

We acknowledge support from the Australian Research Council and NCI National Facility.




\end{document}